\documentclass{an_art}
\usepackage{graphicx}

\newcommand{\be}{\begin{equation}}
\newcommand{\ee}{\end{equation}}
\begin{document}

\begin{titlepage}
\setcounter{page}{000}
\headnote{Astron.~Nachr.~000 (0000) 0, 000--000}

\title {On the time variability of $\gamma$-ray sources: A numerical
analysis of variability indices}

\author{{\sc Diego F. Torres$^{1,2}$, 
M. E. Pessah$^{3}$, Gustavo E. Romero$^{1}$
}
\\
\medskip
{\small $^{1}$ Instituto Argentino de Radioastronom\'{\i}a, C.C.5,
(1894) Villa Elisa, Buenos Aires, Argentina} \\ {\small $^{2}$
Princeton University,
Physics Department, 08544 NJ }\\
{\small $^3$ Facultad de Ciencias Astron\'omicas y
Geof\'{\i}sicas, UNLP, Paseo del Bosque s/n, 1900, La Plata,
Argentina} }
\date{Received ; accepted }
\maketitle

\summary We present a Monte Carlo analysis of the recently
introduced variability indices $\tau$ (Tompkins 1999) and $I$
(Zhang et al. 2000 \& Torres et al. 2001) for $\gamma$-ray
sources. We explore different variability criteria and prove that
these two indices, despite the very different approaches used
to compute them, are statistically correlated (5 to 7$\sigma$).
This conclusion is maintained also for the subset of AGNs and high
latitude ($|b|>10$ deg) sources, whereas the correlation is
lowered for the low latitude ones, where the influence of the
diffuse galactic emission background is strong.
END

\keyw gamma-rays: observations -- variability END
\end{titlepage}

\section{Introduction}

The study of the time variability of $\gamma$-ray sources,
particularly using the Third EGRET Catalog (Hartman 1999), is
currently a very active topic of research. The Third EGRET Catalog
includes observations carried out between April 22, 1991 and
October 3, 1995, and lists 271 point sources. About two thirds of
them have no conclusive counterparts at lower frequencies. Even
worse, 40 of them do not show any positional coincidence (within
the 95\% EGRET contour) with possible $\gamma$-ray emitting
objects known in our galaxy (Romero et al. 1999).

In order to understand the origin of all these unidentified
detections, their variability status is of fundamental importance.
Several known models for $\gamma$-ray sources in our galaxy would
produce non-variable sources during the timescale of observations.
That is the case of pulsars (Thompson 2001) or supernova remnants
in interaction with molecular clouds (Esposito et al. 1996, Combi
et al. 1998, 2001). Alternatively, if some of the sources are
produced by isolated magnetized black holes (Punsly et al. 2000),
microquasars (Paredes et al. 2000), or by stellar winds of early
type stars (Benaglia et al. 2001) one would expect high levels of
flux variability.

Looking at the flux evolution through the different viewing
periods is obviously a first indication of the variability status
of any given source (see for instance Tavani et al. 1997).
However, fluxes are usually the result of only a handful of
incoming photons. A safer way of quantifying the flux evolution
should be devised before obtaining significant results.

\section{Variability indices}

\subsection{The $V$-index for $\gamma$-ray variability}

Three variability indices have been introduced in the literature
so far. The first of them, dubbed $V$, was presented by MacLauglin
et al. (1996), who computed it for the sources contained in the
Second EGRET Catalog. This method was later used, also, for a
short timescale study by Wallace et al. (2000). The basic idea
behind $V$ is to find $\chi ^2$ for the measured fluxes, and to
compute $V=-\log Q$, where $Q$ is the probability of obtaining
such a $\chi ^2$ if the source were constant.  Several critiques
have been mentioned concerning this classification, among them,
that the scheme gets complicated when the fluxes are just upper
limit detections. It can be shown that sources which have upper
limits included in the analysis will have a lower $V$ than that
implied by the data (Tompkins 1999). In addition, a source can
have a large $V$ because of intrinsic reasons --the case we would
be interested in--, or because of small error bars on the flux
measurements. Similarly, a small value of $V$ can imply a constant
flux or big error bars. Each value of $V$ is obtained disregarding
those of a control population. Then, we can have pulsars with very
high values of $V$, or observed AGNs with very low ones. The use
of $V$ to classify the variability of $\gamma$-ray sources seems
not to be very confident.

\subsection{The $\tau$-index for $\gamma$-ray variability}

Tompkins (1999) introduced a new variability criterion which takes
into account not only published EGRET data, contained in the point
source 3EG Catalog, but also unpublished information. In order to
decide the variability index for a given source he used also the
145 marginal sources that were detected but not included in the
final official list, and, all at a time, the detections within 25
deg of the source of interest. The maximum likelihood set of
source fluxes was then re-computed. From these fluxes, a new
statistics measuring the variability was defined as $\tau=\sigma /
\mu$, where $\sigma$ is the standard deviation of the fluxes and
$\mu$ their average value. The strength of this approach lies in
that it takes into account some possible fluctuations from the
background and from neighboring sources, careful sensitivity
corrections throughout EGRET lifetime, and others systematic
errors related either with the equipment itself or with the
processing of the information, in a similar way to that used in
the construction of the 3EG (Hartman et al. 1999, Tompkins 1999).
Details are to be given in Tompkins et al. (2001). The final
result of Tompkins' analysis is a table listing the name of the
EGRET source and three values for $\tau$: a mean, a lower, and an
upper limit (68\% error bars).

\subsection{The $I$-index for $\gamma$-ray variability}

This index was previously used in blazar variability analysis
(Romero et al. 1994) and applied to some of the 3EG sources by
Zhang et al. (2000) \footnote{These authors considered as part of
the control population sources not recognized as pulsars in the
3EG Catalog. See Torres et al. 2001 for a discussion.} and Torres
et al. (2001). The basic idea is to do a direct comparison of the
flux variation of any given source with that shown by pulsars,
which is considered spurious. Then, the $I$-index establishes how
variable a source is with respect to the pulsar population.
Contrary to Tompkins' index, the $I$-scheme uses only the publicly
available data of the 3EG Catalog.

The $I$ index is defined as follows. Firstly, a mean weighted
value for the EGRET flux is computed: \be \left< F \right> =
\left[ \sum_{i=1}^{N_{{\rm vp}}} \frac{F(i)}{\epsilon(i)^2}
\right]\times \left[ \sum_{i=1}^{N_{{\rm vp}}} \frac
1{\epsilon(i)^{2}} \right]^{-1}.\ee $N_{{\rm vp}}$ is the number
of single viewing periods for each $\gamma$-ray source, $F(i)$ is
the observed flux in the $i^{{\rm th}}$-period, whereas
$\epsilon(i)$ is the corresponding error in the observed flux.
These data are taken directly from the 3EG catalog. For those
observations in which the significance ($\sqrt{TS}$ in the EGRET
catalog) is greater than 2$\sigma$, we took the error as
$\epsilon(i) = F(i)/\sqrt{TS}$. For those observations which are
in fact upper bounds on the flux, it is assumed that both $F(i)$
and $\epsilon(i)$ are half the value of the upper bound. Then, the
fluctuation index $\mu$ is defined as: \be \mu =100\times
\sigma_{{\rm sd}}\times \left< F \right>^{-1} .\ee In this
expression, $\sigma_{{\rm sd}}$ is the standard deviation of the
flux measurements, taking into account the previous
considerations.

\begin{figure}
\resizebox{\hsize}{!}{\includegraphics{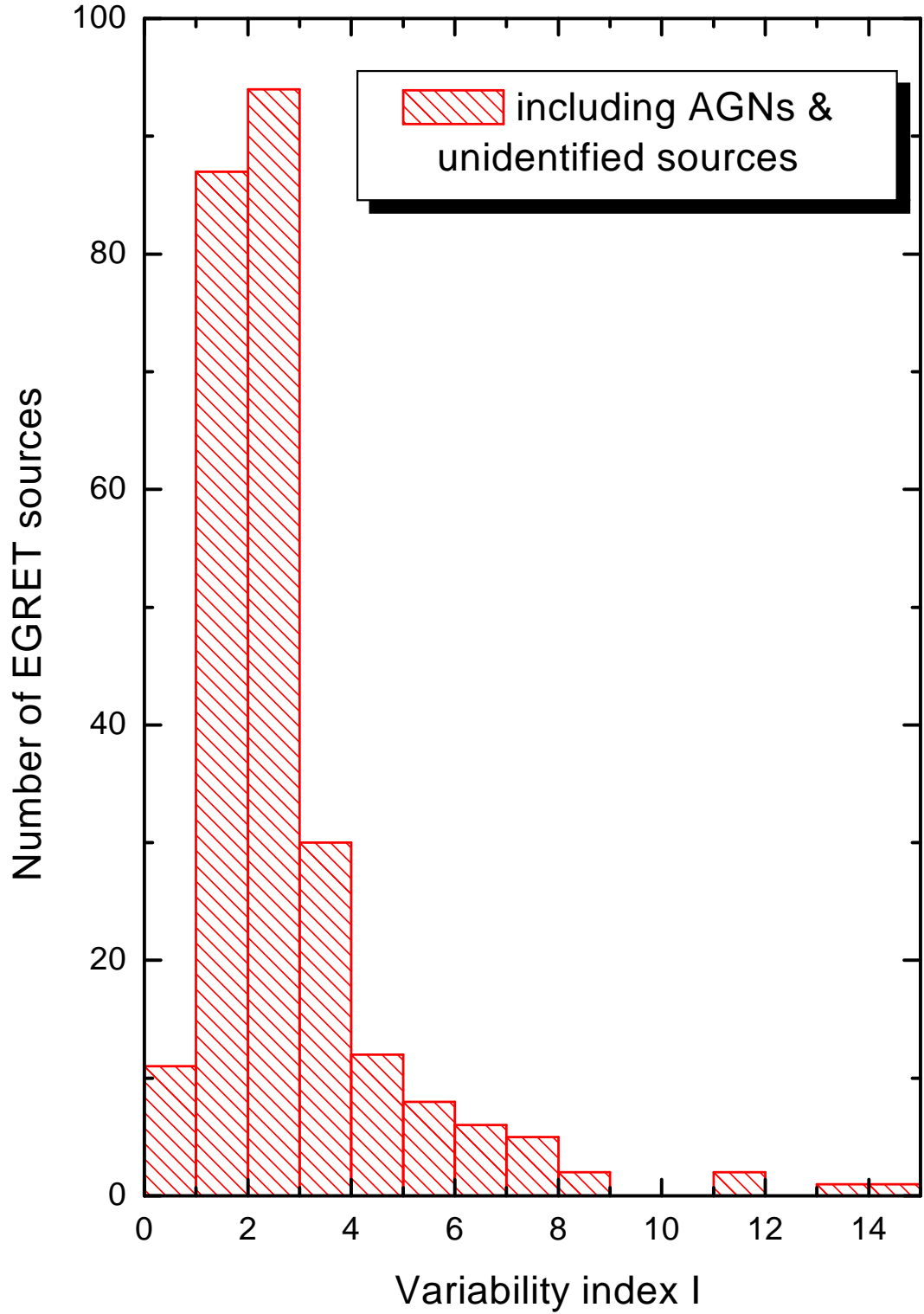}} \vspace{-1.5cm}
\caption{Variability index distribution for 258 $\gamma$-ray
sources in the Third EGRET Catalog, excluding pulsars and six
artifacts related with Vela.}
\end{figure}

\begin{figure}
   \centering
   \includegraphics[angle=-90,width=9cm]{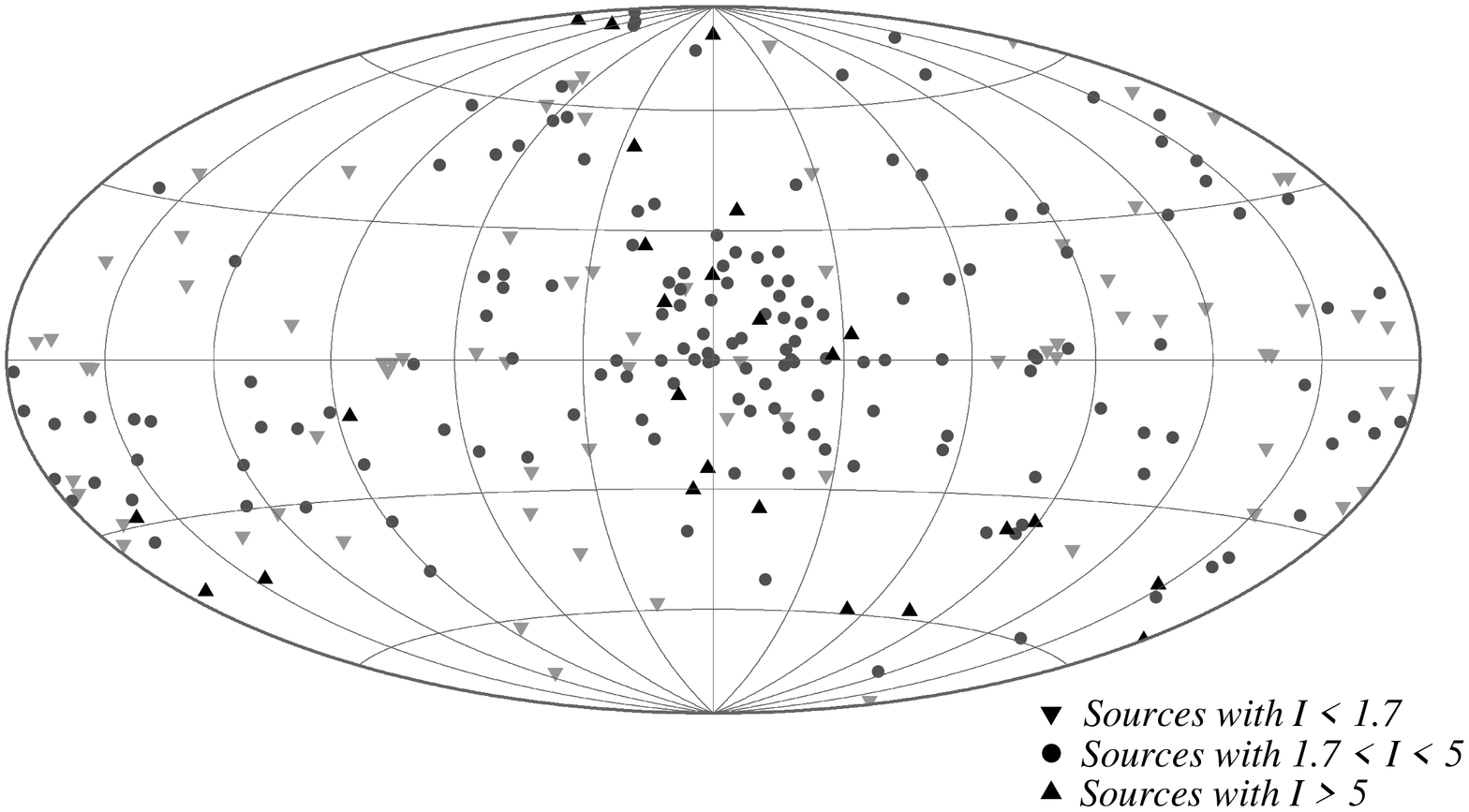}
\caption{Sky variability distribution, under the $I$-scheme, of
258 $\gamma$-ray sources in the Third EGRET Catalog. Up triangles
stand for (25) variable sources, circles for (157) dubious, and a
down triangles for (76) non-variable. Thresholds are as in Table
1.}
\end{figure}

This fluctuation index is also computed for the confirmed
$\gamma$-ray pulsars in the 3EG catalog, assuming the physical
criterion that pulsars are --i.e. by definition-- non-variable
$\gamma$-ray sources. Then, any non-null $\mu$-value for pulsars
is attributed to experimental uncertainty. Finally, the averaged
statistical index of variability, $I$, is given by \be
I=\frac{\mu_{{\rm source}}}{<\mu>_{{\rm pulsars}}}=\frac{\mu_{{\rm
source}}}{26.9}. \ee In Fig. 1 we show the histogram of $I$ for
258 $\gamma$-ray sources in the 3EG Catalog, and in Fig. 2, the
sky distribution.

\section{From variability indices to variability criteria}

\subsection{Plausible criteria for $\tau$}

The index $\tau$ moves from 0 upwards, and it is considered
infinite when it is greater than 10 000. As can be seen from
Tompkins (1999), the thresholds for variability are diffuse. To
have an idea of what a variable source is under the $\tau$-scheme,
Tompkins has separated the $\gamma$-ray sources into different
classes: pulsars, unidentified, AGNs, and sources spatially
coincident with SNRs. He found that $\tau$ can clearly distinguish
between pulsars, whose mean $\tau$-value is 0.1, having the
highest upper limit equal to 0.27, and AGNs, whose mean is 0.9.
The unidentified sources have $\tau$-values pertaining to both
categories. Many sources clearly have a dubious classification;
for instance, 3EG J1339-1419 has a mean $\tau$-value equal to
0.68, but their lower and upper limits are, respectively, 0.17 and
1.70. Then, within the 68\% error bars on $\tau$, this source can
be as variable as an AGN, or as non-variable as a pulsar. This is
an uncomfortably common situation for many sources.

However, if the lower limit on $\tau$ is greater than, say, 1.0,
the source is very likely variable. If the upper limit on $\tau$
is, on the contrary, compatible with the $\tau$-values for
pulsars, we would classify it as non-variable. This encompasses
the spirit of Tompkins' (1999) classification of the most likely
variable and the most likely non-variable sources. The question is
then what the thresholds should be. We have seen that pulsars are
consistent with values of $\tau$ up to 0.27. The deviation for the
mean value of pulsars is $\sim 0.1$. Then, it appears safe to
consider that a source will be likely variable --under the $\tau$
scheme-- when the lower limit on $\tau$ is at least 0.6, $3\sigma$
above the mean value of the $\tau$ upper limit for pulsars.
Equivalently, a source will be considered non-variable when the
upper limit for $\tau$ is below that threshold. Sources not
fulfilling either classification should be considered as dubious.

We can modify these thresholds in a number of ways, but we want
pulsars to represent a non-variable population, and AGNs to be, on
average, a variable one. But even fulfilling these constraints, we
could better use a 2 or 4$\sigma$ level as a safe assumption, or
pretend to artificially move the threshold to $\sim 0.3$, just
above the highest possible level for pulsar variability. We have
explored these assumptions case by case, by means of a computer
code described below, and although we found no statistically
strong variations in the final classification, we did find that a
threshold of 0.5--0.6 is the safer. Known variable sources end up
classified as variable, known or expected non-variable ones also
get their right status.

\subsection{Plausible criteria for $I$}

One possibility for defining a variability criterion for $I$ is
also to consider the error bars for each source: \be \delta I=
\frac{\mu}{ <\mu>_{{\rm pulsars}}^2} \delta <\mu>_{{\rm pulsars}}
\sim 0.5 \;I .\ee Here $\delta <\mu>_{{\rm pulsars}}$ is the
deviation from the $<\mu>_{{\rm pulsars}}$-value. Then, we have
just propagated through $I$ the error in defining the mean value
of the fluctuation index for pulsars. We can then define variable
sources as those fulfilling the constraint \be I - \delta I
> I_p + 3 \sigma ,\ee and non-variable sources as those having \be I + \delta
I  < I_p + 3 \sigma. \ee Here, $I_p=1.0$, is the mean value of $I$
for pulsars, and $\sigma=0.5$, is the deviation in the pulsar
$I$-values. Again, sources not fulfilling neither classification
are to be considered dubious. Then, rephrasing the previous two
equations we get variable sources when $I>5.0$, non-variable
sources when $I<1.7$, and dubious cases for $I$-values in between.
These are very conservative and restrictive constraints, and have
close analogy with the proposed ones for the $\tau$-index.
Particularly, notice that if $I>5.0$ is the threshold for a
variable source, then we are asking for the value of $I$ to be
8$\sigma=8\times 0.5$ times above that of pulsars. Similarly, for
a source to classify as non-variable, its $I$-index should depart
from that of the pulsars in less than 1.4$\sigma$.

This may sound excessive. In addition, why does $3\sigma$ appear in
Eqs. (5) and (6), not 2 or 4? We can as well use the mean value
of $I$ in a direct way, so defining a more straightforward scale.
We can assume a source to be non-variable if its $I$-value is less
than 1.5 (1$\sigma = 1 + 0.5$ above the pulsars), dubious for
$1.5<I<3.5$, and variable for $I>3.5$, 5$\sigma$ above the mean
$I$-index for pulsars. That would be, although less restrictive,
as good a criterion as the previous one. Changing the criteria
will obviously change the variability status of those sources with
values of $\tau$ and $I$ near the boundaries. We then need to
explore all these possibilities in a systematic way before
extracting significant conclusions.

\section{Results of the numerical analysis}

We have written a numerical code that classifies the source
variability, given any chosen criteria, both in the $I$ and the
$\tau$ scheme. In the Internet address provided below we present
complete tables quoting together the $I$ and $\tau$ indices for
each of the sources in the 3EG Catalog. We present in Table 1 the
results for the classification using the above explained criteria:
$\tau$-threshold equal to 0.6, and $I$-thresholds equal to 1.7 and
5.0, respectively. There are 148 detections out of 258 --five
pulsars and six artifacts related with Vela were excluded-- (57\%
of the sources) which classify within the same groups both for $I$
and $\tau$. Can this percentage be obtained randomly?

\begin{table}
\caption{Classification of all 258 sources (excluding pulsars and
six artifacts related with Vela) reported in the Third EGRET
Catalog. Pulsars are non-variable sources in both schemes. The
$\tau$-threshold is equal to 0.6, and the $I$-thresholds are equal
to 1.7 and 5.0, respectively. The row dubbed `Same class' shows
the number of sources that classify within the same group in both
schemes.}
\begin{tabular}{llll}
\hline
Scheme & variable &  dubious & non-variable \\
\hline \hline $I$           &     25   &      157    &      76\\
$\tau$    &  55    &     139      &    64\\
 Same class   & 17       &   95        &  36\\
\hline
\end{tabular}
\end{table}

We have simulated thousands of sets of 258 sources and assigned to
each of them a random variability index $I$. We preserved the
histogram for $I$, i.e. the number of variable, dubious, and
non-variable detections is the same in each of the simulated sets,
but they are assigned to randomly chosen sources. Should we not
preserve the histogram for $I$, we would admit, for instance, a
random case in which all sources are variable, another in which all
are non-variable, etc. This would diminish the random probability
of obtaining the real result in an inappropriate way. What we want
to test is the actual classification which associates a particular
source with a particular value of $I$; this is why, while
maintaining the $I$ distribution, we shuffle the associations.

Not only the percentages of equal classification are important in
order to decide if the two schemes are statistically correlated, but
also the expected random result. For instance, if the thresholds
are chosen such that all sources are non-variable in both schemes,
then the percentage of equal classification would be 100\%. But so
would be the random percentage for each of the simulated sets, and
then there would be no correlation at all.

We found that the expected random result is 104.8$\pm$6.3, i.e.
7$\sigma$ below the real result, implying for it a Poisson
probability equal to 8$\times 10^{-6}$.
We have also used several alternative
plausible thresholds both for $I$ and $\tau$, for instance,
$\tau$-thresholds equal to 0.8, 0.5, and 0.35, with $I$-thresholds
equal to 5.0/2.0, 5.0/1.7, and 3.5/1.5. In all cases, we obtain a
percentage of equal classification above 50\%, the worst random
result (obtained for $\tau$-thresholds equal to 0.35 and
$I$-thresholds equal to 3.5/1.5) being still 5$\sigma$ lower than
the real one. Thus, disregarding the fine grain of the variability
criteria, the two schemes are statistically correlated. We have also
explored what happens if we do not consider those sources having
an average recomputed flux equal to 0.00 within the $\tau$-scheme
(Tompkins 1999). Doing the simulations excluding these sources
produces an even more correlated result.

In Table 2 we show the results for the 67 AGNs. 44 (65\%) of them
have the same classification within both schemes, while we would
expect only 31$\pm$3.0 as a result of chance, 5$\sigma$ lower than
the real result. Again, changing the criteria does not significantly
alter the results (and in most cases actually improves them).

\begin{table} \caption{Classification of confirmed AGNs
reported in the Third EGRET Catalog. Thresholds are as in Table 1.
}
\begin{tabular}{llll}
\hline
Scheme & variable &  dubious & non-variable \\
\hline \hline
$I$           & 10      &    42         & 15\\
$\tau$    &  15      &    40        &  12\\
Same class   & 7       &   30       &    7\\
\hline
\end{tabular}
\end{table}
Table 3 shows the results both for high ($|b|>10$ deg) and low
latitude sources. For the former, the random result is $4\sigma$
below the real one. Changing the criteria to all other plausible
ones we have discussed above enhances the correlation. For the low
latitude sources, the result is $3\sigma$ away from the real one.
Here, changing the criteria to any other of the plausible ones we
mentioned does not generally improve the correlation. The decrease
in statistical correlation between $I$ and $\tau$ at low galactic
latitudes, for the less restrictive criteria, could be reflecting
the uncertainties in the subtraction of the diffuse background
emission of the galaxy (Hunter et al. 1997, Strong et al. 2000).

\begin{table} \caption{Classification of unidentified sources
reported in the Third EGRET Catalog. The upper panel refers to
high latitude sources and the lower one to low latitude
detections. Thresholds are as in Table 1.}
\begin{tabular}{llll}
\hline
Scheme & variable &  dubious & non-variable \\
\hline \hline
$I$           &  12        &  74         & 34\\
$\tau$    &  34        &  61         & 25\\
Same class   & 8       &   39       &   15\\ \hline\hline
$I$           & 3       &   41      &    27\\
$\tau$    &  6        &  38         & 27\\
Same class   & 2       &   26       &   14\\ \hline \hline
\end{tabular}
\end{table}

\section{Discussion and concluding remarks}

The status of a particular source can vary from one scheme to the
other. Then, the joint use of $I$ and $\tau$ can provide a better
idea of the variability status of any given source. Particular
classifications may disagree as a result of completely different
techniques for computing the variability indices. Note that the
dubious classification of any $\tau$ plausible criterion is
applied upon sources we know nothing about. This is not the case
for $I$, since it always provides a scale relative to the mean of the
pulsar fluctuation indices.  Then, it rests on our own
judgement to decide the weight we shall give to a result like
$I=3.0$, but it undoubtedly says that the flux evolution is three
times more variable than the mean flux evolution for pulsars. We
mention that in order to get more reliable results under the
$I$-scheme at low latitudes it seems safer to consider the most
restrictive cutoffs.

It has been noted (R. Hartman, private communication, 2001) that
the weighting used for the definition of $I$ (as done in Zhang et
al. (2000) and Torres et al. (2001)): an inverse square for the
errors in the fluxes, could provide values of $<F>$
unrealistically high. The combined use of unweighted averages, and
the definitions $F \sim \sqrt{TS}/(2+\sqrt{TS})$ and $err(F) \sim
1/(2+\sqrt{TS})$ for the the exposures showing only upper limits,
could even improve the correlation with Tompkins' index. But this
would be another index for quantifying variability, not yet used in
the literature.

\begin{acknowledgements}

This work was partially supported by CONICET, ANPCT (PICT 98 No.
03-04881), and by Fundacion Antorchas, through separate grants to
D.F.T. and G.E.R, and a fellowship to M.E.P.. Tables are available
on-line at http://www.iar.unlp.edu.ar/garra/garra-sdata.html. We
gratefully acknowledge Ms. Paula Turco for her kind help with the
plots. We thank the referee, Dr. R. Hartman, for insightful
remarks that allowed this paper to be improved.
\end{acknowledgements}

\end{document}